\documentclass[pdflatex,sn-mathphys-num]{sn-jnl}

\usepackage{graphicx}%
\usepackage{multirow}%
\usepackage{amsmath,amssymb,amsfonts}%
\usepackage{amsthm}%
\usepackage{mathrsfs}%
\usepackage[title]{appendix}%
\usepackage{xcolor}%
\usepackage{textcomp}%
\usepackage{manyfoot}%
\usepackage{booktabs}%
\usepackage{algorithm}%
\usepackage{algorithmicx}%
\usepackage{algpseudocode}%
\usepackage{listings}%

\begin{document}

\title[Article Title]{A Benchmark for Math Misconceptions: Bridging Gaps in Middle School Algebra with AI-Supported Instruction}

\author*{Nancy Otero} \email{nancy@kitco.design}

\author{Stefania Druga} \email{st3f@uw.edu}

\author{Andrew Lan} \email{andrewlan@cs.umass.edu}

\abstract{This study introduces an evaluation benchmark for middle school algebra to be used in artificial intelligence(AI) based educational platforms. The goal is to support the design of AI systems that can enhance learners' conceptual understanding of algebra by taking into account learners' current level of algebra comprehension. The dataset comprises of 55 algebra misconceptions, common errors and 220 diagnostic examples identified in prior peer-reviewed studies. We provide an example application using GPT-4, observing a range of precision and recall scores depending on the topic and experimental setup reaching 83.9\% when including educators' feedback and restricting it by topic. We found that topics such as ratios and proportions prove as difficult for GPT-4 as they are for students. GPT-4 turbo was used from March 1st to April 25th, 2024.  We included a human assessment of GPT-4 results and feedback from five middle school math educators on the clarity and occurrence of the misconceptions in the dataset and the potential use of AI in conjunction with the dataset. Most educators (80\% or more) indicated that they encounter these misconceptions among their students, suggesting the dataset's relevance to teaching middle school algebra. Despite varied familiarity with AI tools, four out of five educators expressed interest in using the dataset with AI to diagnose students' misconceptions or train teachers. The results emphasize the importance of topic-constrained testing, the need for multimodal approaches, and the relevance of human expertise in gaining practical insights when using AI for human learning. Our dataset and code can be found at: \href{https://github.com/nancyotero-projects/math-misconceptions}{This GitHub repository}, and \href{https://huggingface.co/datasets/nanote/algebra_misconceptions}{this Hugging Face Model}}


\maketitle

\section{Introduction}\label{sec1}

Educational disparities in mathematics persist, particularly for low-income and minoritized students. These students face limited access to resources needed for engaging and supportive learning environments, as well as difficulties in obtaining effective help. \cite{welner2013achievement,ryan2001classroom,puustinen1998help}. Despite efforts to enhance their capacity to provide responsive teaching, personalized mathematics pedagogy remains challenging \cite{dyer2016instructional,hiebert2007preparing,loewenberg2009work}. 

Education technology, such as computer-assisted instruction, adaptive tutors, and massive open online courses (MOOCs), has been viewed as a potential solution to bridge the math achievement gap. However, these technologies often perpetuate existing disparities, with high-income white students reaping more benefits from accessing technology enhanced math-instruction \cite{reich2020remote}. Research on the National Assessment of Educational Progress (NAEP) data from 1996 to 2011 reveals that low-income, non-white children predominantly engaged with technology for repetitive math tasks, while affluent white children used it for more advanced exercises on graphs and problem-solving \cite{wenglinsky1998does}. These persistent socio-economic disparities are widening learning gaps in mathematics education. 

Teachers' understanding of how learners develop mathematical knowledge is crucial for helping low-income and minoritized students understand more advanced topics \cite{hill2018connections}, prior research has shown that addressing math misconceptions has improved learners performance and engagement \cite{parwati2020effectiveness, Ashlock2010ErrorPatterns, li2019problematizing}. Recognizing that learners'math misconceptions are constructive attempts in their process of sense-making allows teachers to leverage students'errors as learning tools to enhance understanding \cite{Ashlock2010ErrorPatterns}. However, accurately measuring student math misconceptions remains challenging despite their predictive value for student learning outcomes \cite{hill2018connections}. 

AI enhanced math-instruction is already being implemented in classrooms with the intention to better serve students across various socio-economic backgrounds \cite{imc2023ai}. But such systems should be capable of adapting to different levels of math conceptual understanding, aiming to provide more equitable access to quality math education \cite{levinson2022conceptions}. Large Language Models (LLMs) have shown promise in generate diverse educational materials \cite{c2023harnessing} including learning objectives, worked examples \cite{jury2024evaluating, prihar2023comparing}, teaching questions, and complete course content \cite{rodriguez2022end, elkins2023useful, wang2022towards, shimmei2023machine, bulathwela2023scalable, doughty2024comparative}. A growing body of research demonstrates the potential for LLMs to enhance learning outcomes in areas such as teacher coaching \cite{pardos2023learning, schmucker2023ruffle, wang2023chatgpt}, student support \cite{dai2023can, xu2023leveraging}, and content development \cite{park2023thinking, xiao2024automation}. In math education LLMs have generated and adapted math problems at various difficulty levels with the goal of enhancing student understanding \cite{jiao2023automatic, norberg2023rewriting}.

Large math datasets with students' interactions exist such as CIMA from \cite{stasaski2020more} or MathDial from \cite{macina2023mathdial} but they are created from synthetic sources. Responses with lower pedagogical quality have result from the use of such datasets \cite{markel2023gpteach, tack2022ai}. Our work uses real interactions between math learning researchers and real students. Additionally, in this study we have included math educators feedback on our dataset for future improvement. 

Our aim is to support algebra educators in better identifying misconceptions and errors (MAEs) in the classroom. This focus is particularly important given that passing 6th-grade math is a strong predictor of high school graduation \cite{balfanz2007preventing}. The primary research question addressed in this study is: How can AI solutions help educators identify students' misconceptions? We propose a three-part solution:
\begin{enumerate}
    \item The creation of a comprehensive algebra misconception benchmark for 4rd-8th grade
    \item An evaluation of LLMs' accuracy in diagnosing each of the 55 misconception using the examples of the misconceptions from our dataset.
    \item Algebra educators' feedback on the benchmark's potential use and on the benchmark's evaluation results.
\end{enumerate}

Our study bridges math misconception research with AI-driven educational tools by creating a comprehensive algebra misconceptions dataset. We evaluate LLMs' ability to diagnose these misconceptions, achieving 83.91\% accuracy in detecting misconceptions from question/answer pairs. Our results highlight the complexity of proportional reasoning and areas for improvement in LLM performance. Notably, 80\% of surveyed teachers found our examples clear, recognized these misconceptions in their classrooms, and expressed interest in using the dataset for diagnosis. This work provides a valuable resource for researchers, technologist and educators aiming to improve math learning outcomes and suggests that LLMs can become a valuable tool in supporting math education.

This dataset aims to bridge the gap between math learning research and the practical needs of educators, particularly for low-income and minoritized learners. Future research can explore the benchmark's effectiveness in real educational contexts or using other LLMs and prompting methodologies. Its applications for AI-driven educational tools can include adapting interactions to learners' specific misconceptions, informing teachers about classroom learning gaps, matching misconceptions with the understanding needed to overcome them, and supporting curriculum development that proactively addresses known difficult areas in math.

\section{Related Work}\label{sec2}
\subsection{Common Errors and Algebra Misconceptions in Middle School}\label{sec2.1}

Teachers' understanding of how learners develop mathematical knowledge is crucial for helping low-income and minoritized students understand more advanced topics \cite{settles2020machine}. This understanding is particularly important because mistakes and misconceptions, far from being mere obstacles, are important learning opportunities \cite{boaler2013ability}. Recognizing that learners' math misconceptions are constructive attempts in their process of sense-making allows teachers to leverage students' errors as powerful learning tools \cite{moore2022assessing}. This approach not only facilitates meaningful student learning and retention \cite{stefanich1992analysis, wilcox1997implementing, riccomini2005identification, stein2007modular, schnepper2013analysis} but also improves learners' performance and engagement \cite{hocky2022natural, moore2022assessing, walsh2022lesson}. In the context of math education, scaffolding based on closing  knowledge gaps has been shown to be a highly effective instructional approach, with meta-analyses reporting significant effect sizes on student outcomes \cite{hattie2008visible, zuo2024relationship}.
Effective remediation involves educators engaging deeply with mathematical details in student responses, including errors, which not only enhances understanding but also fosters strong teacher-student relationships, boosting student motivation \cite{wentzel1997student, wentzel2022does, pianta2003improving, robinson2022systematic, easley1975teaching, brown1978diagnostic, carpenter1999children, carpenter2003thinking, lester2007second, loewenberg2009work}. Despite the clear benefits and predictive value of addressing misconceptions for student learning outcomes, accurately measuring and responding to these misconceptions in real-time remains a significant challenge in educational research and practice, particularly for novice teachers who struggle to identify misconceptions and common errors (MaEs) \cite{settles2020machine}. Our work looks to create a list of researched MaEs to facilitate their recognition in the classroom and to be incorporated into math learning platforms. 

Comprehensive lists of misconceptions exist for some subjects, such as the Misconception-Oriented Standards-based Assessment Resources for Teachers (MOSART) \cite{mosart2024h} for science. MOSART supports science educators in knowing the science misconceptions present in their classrooms, when educators know their students misconceptions is easier for them to use the decades in research on how to overcome those misconception \cite{doyle2020professional,sadler2013influence}. A comparable resource for mathematics is lacking. Closing this gap is fundamental for leveraging research on both detecting and overcoming math misconceptions. Educators need to understand the specific misconceptions present in their classrooms to effectively apply practices in overcoming math misconceptions. Some efforts have been made to compile math misconceptions, such as Ashlock's (2010) \cite{Ashlock2010ErrorPatterns} work on arithmetic and Bush's (2011) \cite{bush2011analyzing} research on pre-algebra. However, a comprehensive list that can be used for assessment and integrated with Large Language Models (LLMs) is still needed.

\subsection{AI and Math Instruction}\label{sec2.2}
The integration of artificial intelligence (AI) into mathematics education has been rapidly evolving. Digital math learning platforms, used by millions of students daily, have shown positive results. Computer-based tutors that detect student errors and provide immediate feedback have demonstrated increased student success \cite{mckendree1990effective, koedinger2007exploring, kantack2022instructive}. These technologies are now candidates for AI integration, particularly through Large Language Models (LLMs).
LLMs have shown significant potential in math education, with classroom implementations already occurring \cite{imc2023ai}. They enhance automated scoring of math questions, achieving human-like agreement in some cases \cite{c2023harnessing, jury2024evaluating, elkins2023useful, pardos2023learning, morris2024automated}. LLMs generate diverse educational materials, including learning objectives, worked examples \cite{jury2024evaluating, prihar2023comparing}, teaching questions, course content \cite{rodriguez2022end, elkins2023useful, wang2022towards, shimmei2023machine, bulathwela2023scalable, doughty2024comparative}, and adapted math problems at various difficulty levels \cite{jiao2023automatic, norberg2023rewriting}. Research demonstrates their potential to improve learning outcomes through teacher coaching \cite{pardos2023learning, schmucker2023ruffle, wang2023chatgpt}, student support \cite{dai2023can, xu2023leveraging}, and potentially doubling learning outcomes in AI-assisted tutoring \cite{wang2024bridging}.
LLMs have shown promise in math content generation \cite{park2023thinking, xiao2024automation}, including automated distractor and feedback generation for multiple-choice questions \cite{lee2024math}. They've outperformed existing approaches in algebra error classification \cite{mcnichols2023algebra}. However, LLMs still struggle to identify specific math misconceptions \cite{Liu2023Novice, gorgun2023enhancing}, generate answers demonstrating misconceptions \cite{Liu2023Novice}, or explain incorrect answers. with new research on automating misconception identification in student explanations showing potential \cite{dai2023can}. There's a need for these systems to adapt to different levels of math understanding for equitable education access \cite{levinson2022conceptions}.
Large math datasets with student interactions, like CIMA \cite{stasaski2020more} and MathDial \cite{macina2023mathdial}, are crucial for training AI systems. However, synthetic datasets can result in lower pedagogical quality responses \cite{markel2023gpteach, tack2022ai}. The Bridges dataset \cite{wang2024bridging}, with 700 real math tutoring conversations, shows potential for embedding expert thought processes in LLMs, improving responses by 76\% when used with GPT-4.
Our work aims to contribute datasets of real interactions between math learning researchers and students, incorporating math educators' feedback when used to enhance for future improvements. This addresses the need for more authentic and pedagogically robust AI used data in mathematics education.

\section{Method}\label{sec3}

We developed a dataset comprising 55 distinct algebra misconceptions and common errors (MaEs), each accompanied by four diagnostic examples, resulting in a total of 220 examples. The dataset was created using a snowball sampling method \cite{wohlin2022successful}, analyzing 145 peer-reviewed journal manuscripts, conference proceedings and papers, books, and dissertations. The misconceptions are characterized by dimensions such as algebra topic and description of the misconception. The topics covered are based on \cite{welder2006prerequisite}, \cite{welder2007preservice}, \cite{welder2012improving} categorization \cite{bush2011analyzing, bush2013prerequisite, otero2024mathmisconceptions} and include algebra-related MaEs observed during 4th or 5th grade, as these can impact students' understanding of middle school algebra.

Our dataset is specifically structured to be compatible with Large Language Models (LLMs), enabling them to recognize patterns in student errors across various algebra topics and match student responses to known misconceptions more accurately.
We focus on algebra as it is a critical prerequisite for success in high school, postsecondary education, and STEM careers \cite{capraro2006algebraic, erbas2005predicting}.

To evaluate this benchmark's characteristics and applications, we conducted a two-part study: 
\begin{enumerate}
    \item \textbf{LLM Performance Evaluation}: We measure GPT4-turbo\footnote{Accessed from March 1st to April 25th, 2024.} ability to identify misconceptions using our benchmark with two standard metrics \cite{sajjadi2018assessing} precision (how many of the misconceptions identified by the LLM were correct) and recall (LLM's ability to find all relevant instances of the misconception). 
    \item \textbf{Educator Feedback}: We invited feedback from algebra middle school educators on the LLM's errors to understand where and why the model might be struggling. We also surveyed these educators on the clarity of the misconceptions in our benchmark and how often they encounter these misconceptions in their classrooms. 
\end{enumerate}

\subsection{Dataset Creation}\label{se3.1}

The dataset creation process began with a comprehensive search on Google Scholar for peer-reviewed conference papers and journal articles on algebra misconceptions. We then followed references and citations from these initial sources to expand our research collection. For each unique MaE identified in the literature, we searched for examples of how researchers diagnosed the misconception, aiming to collect four math problems with incorrect answers for each MaE. This approach ensured multiple ways to diagnose each misconception and confirmed that other researchers had identified the same MaE. We analyzed a total of 145 sources, with 56\% of the research being conducted in the US. When we found more than four examples for a MaE, we prioritized those from research with diverse and underserved populations. The scope of our research was considered complete when we could no longer find at least four examples to diagnose a newly discovered MaE, or when the MaE was too general to diagnose using only one question and answer.
Within pre-algebra and algebra instruction, we address two different learning processes: misconceptions and errors. Errors are defined as learners' computational errors indicating that they have not mastered using a rule. Overcoming math errors is crucial because they can persist for years, affecting the learner's ability to build more complex math skills \cite{wenglinsky1998does}. For instance, Bush (2011) \cite{bush2011analyzing} found that 16.5\% of middle school learners made basic computational errors with whole numbers when answering algebra questions. An example of a computational error is consistently making the sum of two negatives a positive, such as -6 – 3 = 9 or -5 – 4 = 9 \cite{Ashlock2010ErrorPatterns}, or when students struggle with plotting points, reversing the x- and y- coordinates \cite{bush2011analyzing}. The reasons why students make computational errors can be multifaceted, including a lack of understanding of the rule, a misconception about the rule, an underdeveloped understanding of a particular underlying concept, an error while learning the rule, or a mistake in applying the rule \cite{Ashlock2010ErrorPatterns}.
On the other hand, misconceptions are mistakes in conceptual understanding that can affect all applications of an individual concept \cite{skemp1976relational}. For example, a single misconception about proportional relationships (part/whole, part/part, whole/part) can lead to problems in identifying proportions in fraction figures and the failure to realize that all parts must be of equal size \cite{van2019primary}. Thus, the denominator of the fraction is associated with the total number of parts regardless of their size. This misunderstanding can then extend to more advanced concepts, such as the slope of a function. The slope represents the rate of change between two variables and is fundamentally a type of proportional relationship, a ratio. If a learner struggles with basic proportional reasoning, they may have difficulty grasping that slope represents how much the y-value changes for each unit change in the x-value, making understanding linear functions challenging \cite{van2019primary}.

The misconceptions in the dataset are organized according to the following topics, as outlined by Welder:

\begin{enumerate}
    \item \textbf{Number sense:} A good intuition about numbers and their relationships \cite{howden1989teaching}. (MaE01 - MaE05)
    \item \textbf{Number operations:} The ability to subtract integers, represent and explain fractions and fraction operations, decimal representations, prime numbers, and the order of operations \cite{welder2007preservice}. (MaE06 - MaE22)
    \item \textbf{Ratios and proportional thinking:} The ability to reason proportionally, to understand ratio concepts, and to use ratio reasoning to solve problems \cite{ojose2015proportional}. (MaE23 - MaE28)
    \item \textbf{Properties of numbers and operations:} Understanding of the commutative, associative, and distributive properties; being capable of performing algebraic manipulations and understanding the order of operations \cite{bush2011analyzing}. (MaE31 - MaE34)
    \item \textbf{Patterns, relationships, and functions:} The ability to represent, analyze, and generalize a variety of patterns with tables, graphs, words, and symbolic rules \cite{bush2011analyzing}. (MaE35 - MaE42)
    \item \textbf{Algebraic representations:} Relating and comparing different forms of representation for a relationship, exploring relationships between symbolic expressions and graphs, using symbolic algebra to represent situations, generating equivalent forms for algebraic expressions, and solving linear equations \cite{bush2011analyzing}. (MaE43 - MaE44)
    \item \textbf{Variables, expressions, and operations:} The ability to see structure in expressions, perform arithmetic with polynomials and rational expressions, create equations, and reason with equations and inequalities \cite{bush2011analyzing}. (MaE45 - MaE48)
    \item \textbf{Equations and inequalities:} Create equations, and reason with equations and inequalities, understand the connections between proportional relationships, lines, and linear equations, define, evaluate, and compare functions and use them to model relationships between quantities \cite{bush2011analyzing}. (MaE49 - MaE55)
\end{enumerate} 

To enhance the dataset's utility and consistency, we implemented several key formatting decisions. The correct answer to each question was included to provide a baseline for LLM knowledge. In some instances, we reformatted questions and answers from their original form to ensure a more homogeneous presentation across different examples of each MaE. Additionally, we removed researchers' explanations from the students' answers to prevent this feature from influencing the accuracy and recall experiments.

\section{LLM Experiments}\label{sec4}

\begin{figure}[H]
    \centering
    \includegraphics[width=0.9\linewidth]{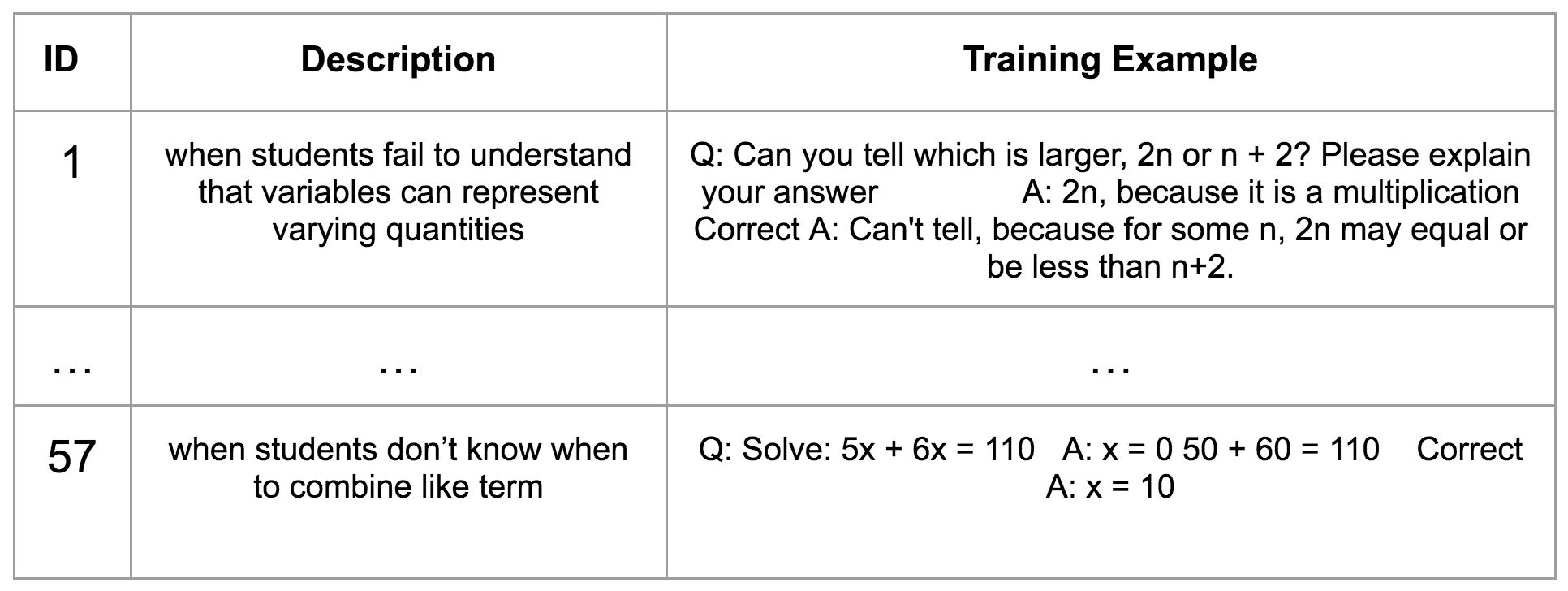}
    \caption{Example of a training set}
    \label{trainingset}
\end{figure}

\begin{figure}[H]
    \centering
    \includegraphics[width=0.4\linewidth]{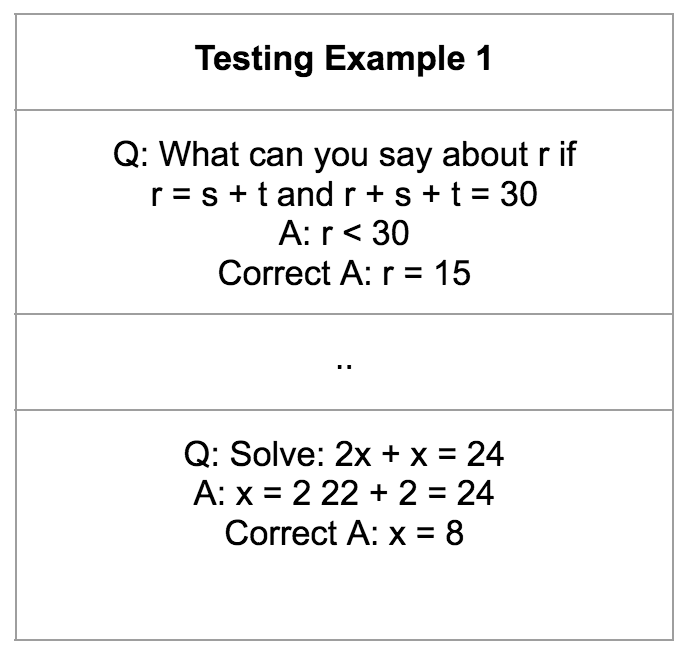}
    \caption{Example of a test set}
    \label{testset}
\end{figure}

We employed a repeated random subsampling validation approach to evaluate the performance of the gpt4-turbo model \cite{open2303gpt} in predicting misconceptions and to investigate the relationships between misconceptions' examples. In each iteration, one example out of the four per MaE was randomly selected as the training set, and another example per MaE was randomly selected to form the test set. The model was trained using in-context learning on the training set (see Fig. \ref{trainingset}) and evaluated on the test set (see Fig. \ref{testset}); we chose in-context learning since this method was previously used to evaluate LLMs performance in tasks related to algebra misconceptions in educational settings \cite{Liu2023Novice}. 

\textbf{The prompt for the experiments was as follows}
Prompt: 
"You are an expert tutor on middle school math with years of experience understanding students' most common math mistakes.
You have identified a set of common mistakes called Misconceptions, and you use them to diagnose student's answers to math questions.
You have also developed a labeled dataset of question items, and diagnosed them with the appropriate misconception id.
Using the set of misconceptions and the labeled dataset, your task today is to take some items of unlabeled data and provide a diagnosis for each unlabeled item.

Here is the list of misconceptions together with a brief description:
{description}
Here is  dataset of question items already labeled with a diagnosis:
\{train\_example\}
Here is the unlabeled dataset of question items. For each question item below, give a diagnosis.
\{items\_test\}"

See Appendix for the complete prompt used. 

This process was repeated for 100 iterations to ensure the robustness and generalizability of the results. As we used gpt4-turbo, our experiments used only text. When an image was part of the question or answer, we included the description of each image (see Appendix \ref{A1}). The gpt4-turbo predicted misconception using as training an example of a misconception and the actual misconceptions were recorded on each iteration. 

This approach allowed us to investigate the potential for multiple misconceptions to match a single wrong student's answer and provides practical insights on the use of each example as diagnostic tools. A student's incorrect response may result from multiple underlying misconceptions in real-world scenarios. By examining the model's predictions across different iterations and comparing them to the actual misconceptions, we can gain insights into the relationships between examples and the co-occurrence of misconceptions. These insights can be valuable in designing targeted interventions to address specific misconceptions.

The results from all iterations were combined and analyzed to calculate precision and recall metrics at both the example and misconception levels. In addition, we examined the frequency of predicted and actual misconceptions for each example and the co-occurrence of misconceptions within the same example. This analysis provides a more nuanced understanding of the model's performance and sheds light on the complex relationships between examples and misconceptions.

The second experiment was similar to the first. However, the examples for the testing dataset were randomly selected from a list of examples belonging to the same topic as the training MaE (e.g., Number sense, Equations, and inequalities). As in the first experiment, when an image was part of the question or answer, we included the description of each image. The process was also repeated 100 times. Precision and recall were calculated as in Experiment 1 (Fig. \ref{recallperMaE}).

\section{Structure Interview and Survey}\label{sec5}

We conducted an online survey with five teachers and four from those five teachers participated on an online structure-interview \cite{cohen2017action}. The goal was to gather feedback from algebra middle school educators that server low-income and minoritize students on the clarity and occurrence of the misconceptions in our dataset and the potential use of AI in conjunction with the dataset. The survey consisted of three sections:
\begin{enumerate}
    \item This section provides information about teachers' experiences, such as their years of teaching, the grade levels they teach or have taught, and their familiarity with algebra misconceptions.
    \item This section included questions regarding the clarity of examples of misconceptions and their occurrence in the classroom. Eight misconceptions with one example were randomly selected from each topic on the data set. Educators were asked to rate the clarity of each of the eight misconception descriptions and examples provided and indicate if they have encountered these misconceptions in their classrooms.
    \item This section included questions about their current use of AI and the potential use of the dataset with AI in teaching practice.
\end{enumerate}

We sought out educators who work with students from underrepresented groups and underserved communities. Five educators with 4 to 10+ years of teaching experience in grades 6 through 10 at public schools participated in our survey. Four of these educators currently teach in schools serving underserved populations, including one who works in a youth correctional facility. We do not have information about the teaching context of the fifth educator. Regarding the clarity and occurrence of the misconceptions presented, most educators (80 \% or more) indicated that they encounter these misconceptions among their students. Four of the five educators explained during the interview that they have not seen misconceptions on some topics because they do not teach that topic in their grade. This high level of agreement suggests that the misconceptions included in our dataset are indeed prevalent and relevant to teaching middle school algebra.

In terms of familiarity with AI tools, three out of five educators reported being unfamiliar with educational AI tools, while one was somewhat familiar and one was very familiar. Despite this varied familiarity, four out of five educators expressed interest in using the dataset with AI tools to diagnose students' misconceptions or train teachers about misconceptions. One educator suggested using the dataset for a chatbot or an online test, while another mentioned using it for a diagnostic test.

We structured the interview to focused on these questions:
\begin{enumerate}
    \item Which other misconceptions do you see often in your classroom?
    \item How can we improve our misconception's examples?
    \item Would you be interested in using out dataset in your classroom? If so, how would you like to use our dataset in your classroom?
    \item How can AI support the use of our dataset in your classroom?
\end{enumerate}

For the first question, one teacher expressed the need to include misconceptions from other topics, such as geometry, three teachers mentioned misconceptions that are part of the dataset but were not shown in the online survey, and one teacher mentioned a new misconception we did not find in our search. Two teachers mentioned that the misconception about reversing the x- and y-coordinates was common some years ago in their classroom but is no longer expected since students use platforms such as Minecraft and Scratch. 

For the second question, all teachers recommended improvements to the format of the misconception examples. We anticipated this feedback due to limitations of our survey platform, which did not allow for clear and standardized presentation of mathematical expressions. Three teachers found the same two answers out of the eight misconception examples "not very clear." In both examples, we removed the researcher's comments on the student's answer because having those comments made the LLM identify the correct MaE more easily. Teachers' comments suggest that we can improve out dataset by including educators' feedback.

All four teachers wanted to use our dataset to diagnose misconceptions. Three teachers mentioned they would like to read it to learn more about students' misconceptions. One teacher said it was very interesting that our answers showed students' processes because most of the time, she could only see the wrong answer through an online platform and did not know the students' processes.

During the last question, the four teachers wanted to use AI to help them diagnose the misconceptions. Three wanted a test that could use AI to find misconceptions; one preferred a chatbot that could ask students about their math thinking process. Another teacher answered: "The use of AI to teach mathematics to children is a good option, but I believe that before using it, children should know about the basic mistakes they make while attempting mathematics problems... Children have to be made to understand that success is achieved only by repeated efforts, and we learn from mistakes. We have to make children understand how to learn from mistakes." This comment suggests a possible use of the dataset as part of a process where students learn from theirs or others mistakes which is an application aligned with the intention for creating the dataset. 

\section{Results}\label{sec6}

\begin{figure}[H]
    \centering
    \includegraphics[width=0.75\linewidth]{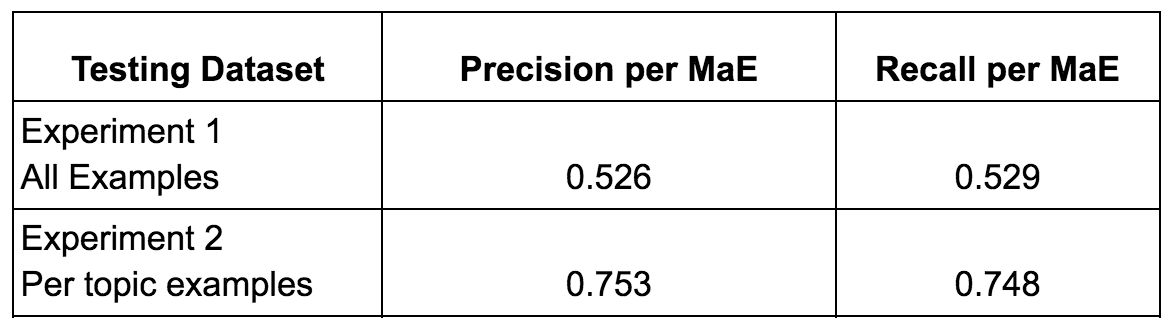}
    \caption{Results for precision and recall per MaE}
    \label{recallperMaE}
\end{figure}

The results of the two experiments demonstrate the model's performance in predicting misconceptions using the dataset. In the first experiment, where testing examples were randomly selected from the entire dataset, the model achieved a precision of 0.526 and a recall of 0.529 per MaE. In the second experiment, where testing examples were randomly selected within each topic, the model achieved a higher precision of 0.753 and a recall of 0.748 per MaE (see Fig. \ref{recallperMaE}).
\begin{figure}[H]
    \centering
    \includegraphics[width=0.75\linewidth]{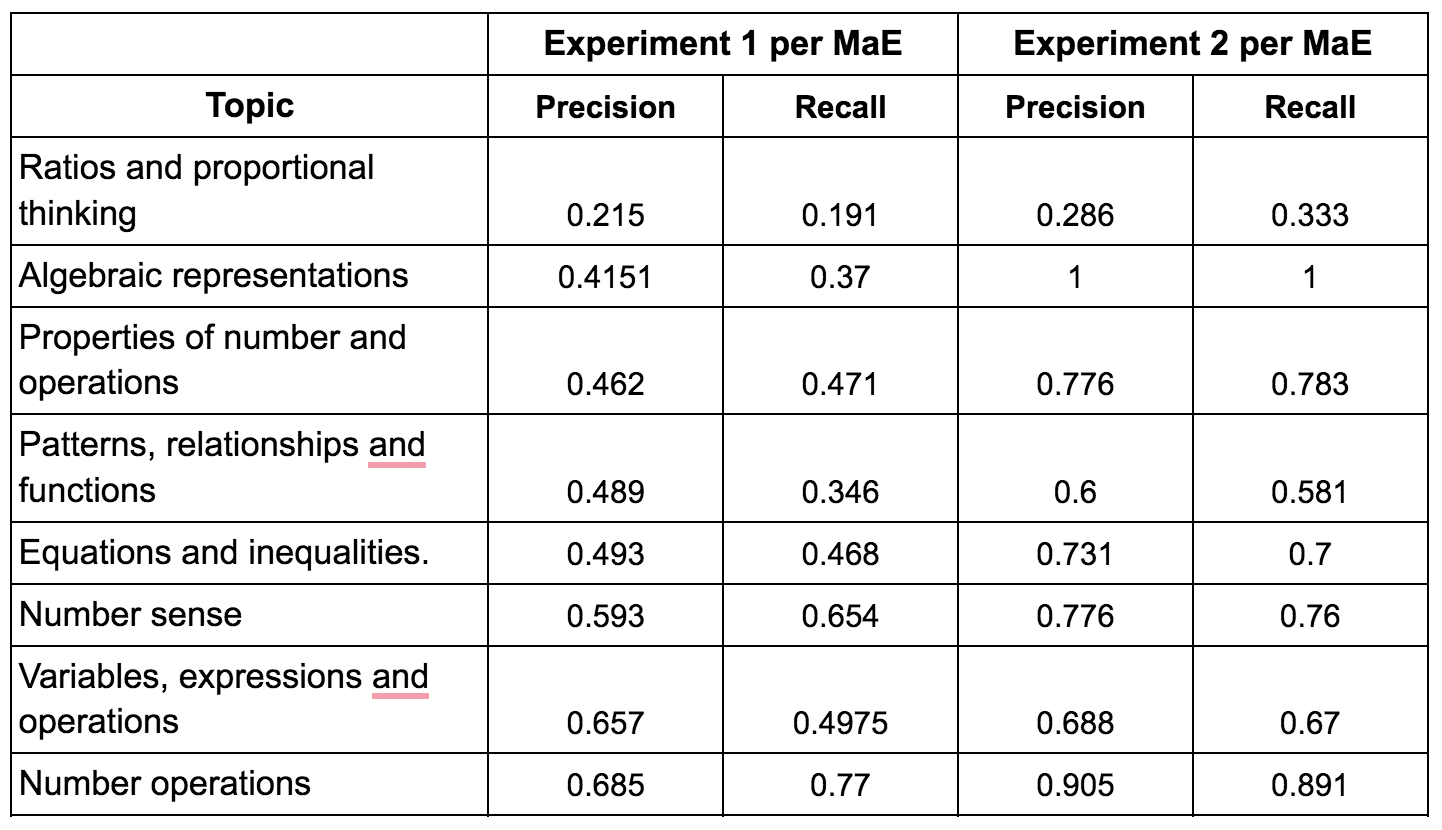}
    \caption{Results for precision and recall per MaE per topic}
    \label{perMaEpertopic}
\end{figure}
\begin{figure}[H]
    \centering
    \includegraphics[width=1\linewidth]{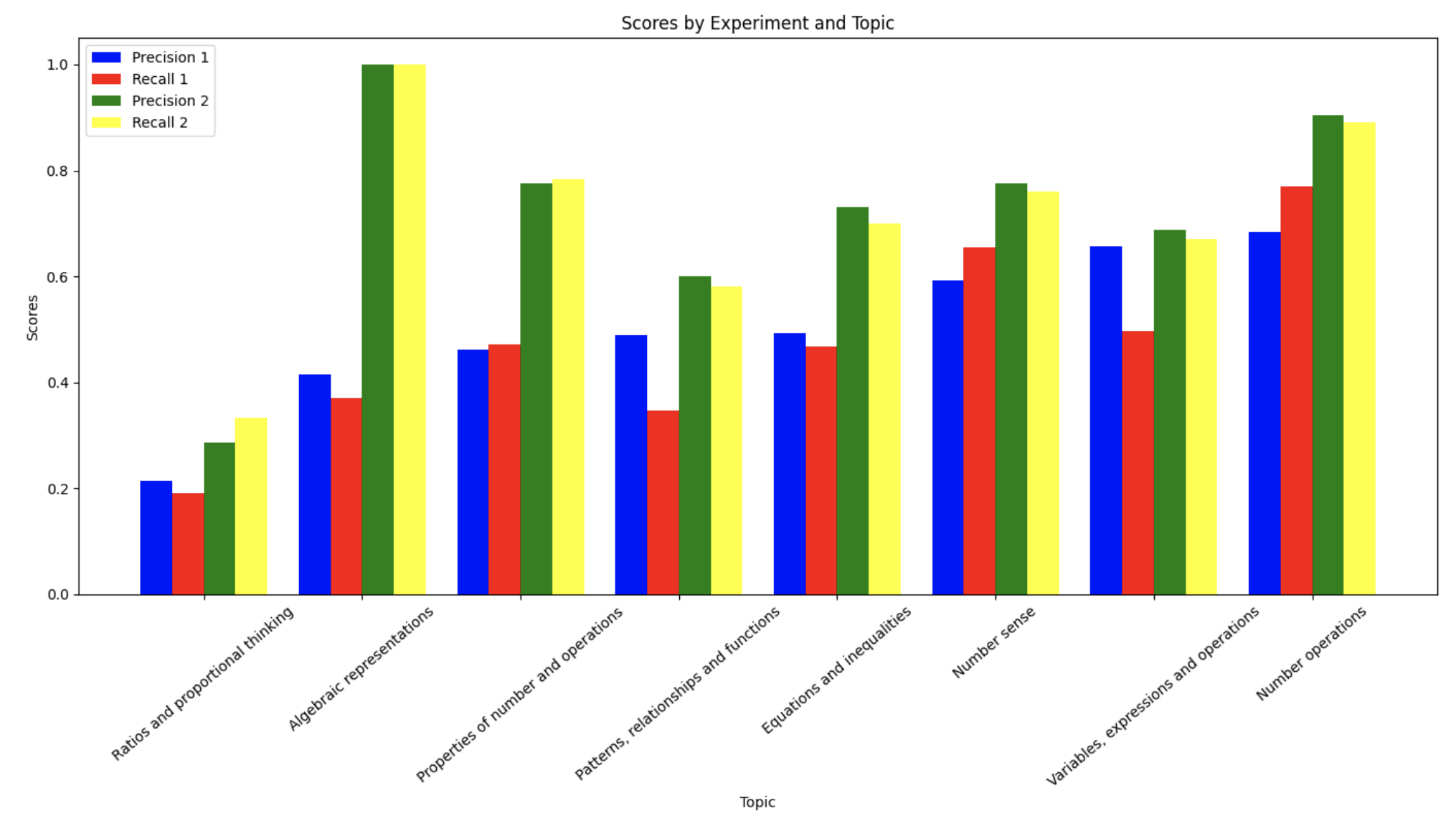}
    \caption{Scores by Experiment and Topic}
    \label{bargraph}
\end{figure}

The table presenting the results of experiments per topic provides a more granular view of the model's performance. The precision and recall scores vary considerably across topics; in Experiment 1, where testing examples were randomly selected from the entire dataset, the model achieved the highest precision (0.685) and recall (0.77) for the "Number operations" topic, indicating its proficiency in identifying misconceptions related to this area. Conversely, the model struggled the most with the "Ratios and proportional thinking" topic, yielding the lowest precision (0.215) and recall (0.191) scores (see Fig. \ref{bargraph}). 

In Experiment 2, where testing examples were randomly selected within each topic, the model's performance improved across all topics, with the most significant gains observed in the "Algebraic representations" topic, achieving perfect precision and recall scores of 1. However, even with topic-constrained testing, the model still exhibited relatively lower performance in the "Ratios and proportional thinking" topic, with precision and recall scores of 0.286 and 0.333, respectively (Fig. \ref{bargraph}). 

These results underscore the inherent complexity of proportional reasoning, a skill that develops over time and often proves challenging for students to grasp \cite{lamon2007rational}, and the need for further research to accurately improve the model's ability to identify misconceptions in this area.

The confusion matrix (CM) presented in figure \ref{confusionmatrix} provides a detailed view of the target versus predicted MaEs in experiment 1. This confusion matrix displays the model's performance in predicting 55 distinct algebra misconceptions and errors (MaEs). The x-axis represents the predicted MaEs, while the y-axis shows the target (actual) MaEs. Darker shades indicate higher frequencies of predictions. The diagonal elements represent correct predictions, while off-diagonal elements show misclassifications. Notable patterns include stronger performance in certain MaEs (e.g., MaE05 and MaE20) and consistent misclassifications between specific pairs (e.g., MaE07 and MaE11, MaE44 and MaE49). The matrix reveals hierarchical relationships between misconceptions, such as those from Patterns, Relationships, and Functions (PRF) being classified as Ratios and Proportions (RP), but not vice versa. This visualization helps identify areas where the model excels and where it struggles, providing insights for targeted improvements in misconception identification.

We also explored the hierarchical structure of misconceptions to understand how one misconception might lead to another. This understanding can guide educators and online platforms in addressing learners' mathematical errors (MaEs) more effectively once identified. It can also help anticipate which MaEs to address before teaching a particular topic. By gaining insights into these relationships between misconceptions, educators can develop more targeted strategies for addressing students' knowledge gaps and prevent new misconceptions from forming due to existing ones. To investigate these relationships, we conducted further analysis of the Large Language Models' (LLMs) mistakes.

There are notable off-diagonal elements, suggesting that the model struggles to distinguish between specific pairs of MaEs. For instance, the matrix shows a relatively high frequency of misclassification between MaE07: \textit{"when students simplify just one of the terms in a fraction, either the numerator or the denominator"} and MaE11: \textit{"when students subtract mixed numbers incorrectly, avoiding regrouping and just subtracting the smaller from the larger number"} and between MaE44: \textit{"when students struggle to grasp the concept of independent and dependent variables"} and MaE49: \textit{"when students find it challenging to comprehend the various meanings and applications of variables"}, indicating that the model often confuses these specific misconceptions.

To further investigate the model's performance, we reviewed the MaEs that were incorrectly classified. We noticed that in some of them, besides the MaE diagnosed by researchers, more than one MaE could be detected with the researcher's example.
We then invited two experienced algebra educators, that did not participate in our structured interview, to review the MaE labeling errors made by the model. Through our analysis and the discussion with the algebra teachers, we classified the errors into the following categories:
\begin{enumerate}
    \item Agreed with AI
    \begin{enumerate}
        \item Because more than one of the MaEs could be matched to that example (Appendix \ref{A3} Figures \ref{Table5.1} \& \ref{Table5.2})
        \item A MaE is a subset of another MaE (Appendix \ref{A3} Fig. \ref{Table6})
    \end{enumerate}
    \item Picture needed for correct classification. The question asked to educators was: Will they identify the correct MaE by only reading the picture description, like the LLM? (Appendix \ref{A3} Fig. \ref{Table7})
    \item The AI indeed makes a mistake (Appendix \ref{A3} Fig. \ref{Table8})
\end{enumerate}
\begin{figure}[H]
    \centering
    \includegraphics[width=0.75\linewidth]{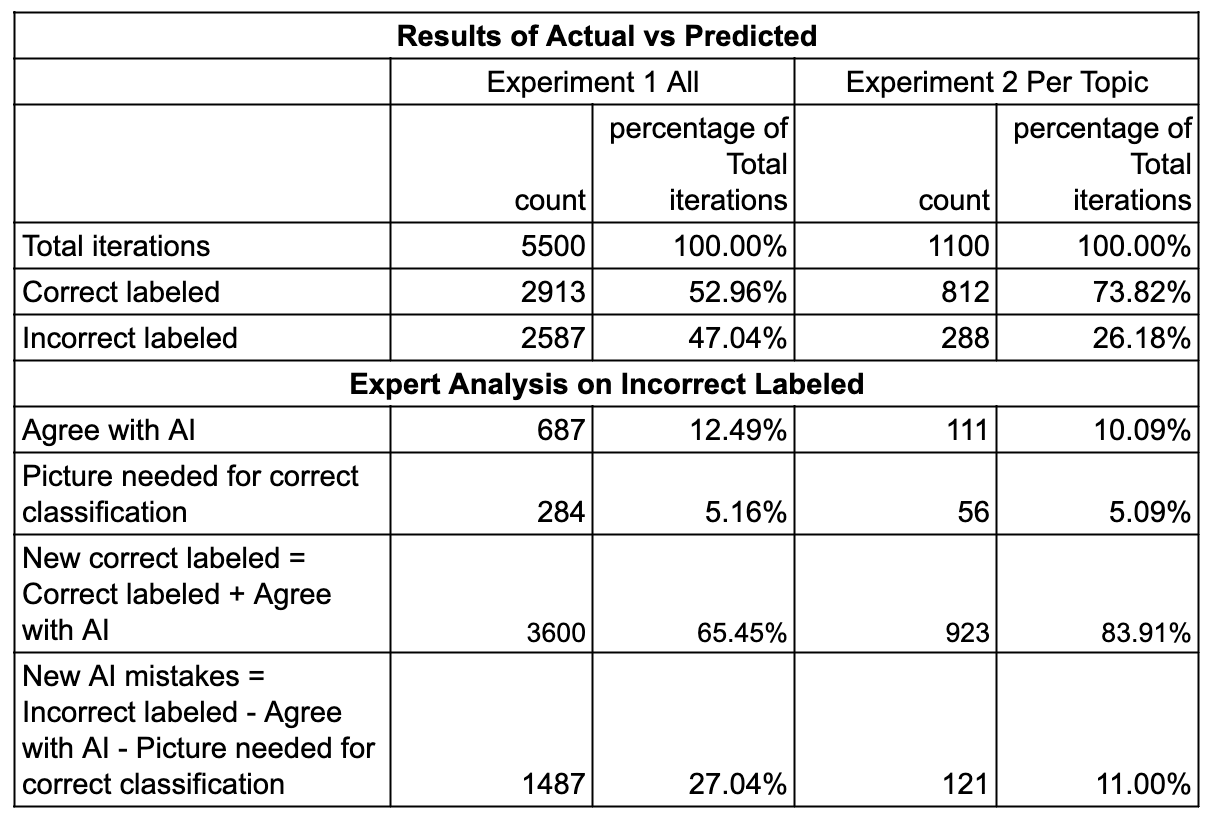}
    \caption{Results of Actual vs Predicted}
    \label{actualvspred}
\end{figure}

The two expert educators agreed on 90.91\% of their answers. A session that included both educators was held to re-rate the pair of MaEs they disagreed upon until just one pair disagreed. 

The table presents the expert analysis results on the incorrectly labeled MaEs by the AI model in both Experiment 1 and Experiment 2. In Experiment 1, where all examples were considered, there were 5,500 iterations. The AI model correctly labeled 2,913 (52.96\%) of the iterations and incorrectly labeled 2,587 (47.04\%). Out of the incorrectly labeled iterations, the experts determined that in 1,487 (27.04\%) cases, the AI was mistaken, while in 687 (12.49\%) cases, the experts agreed with the AI's labeling. In 284 (5.16\%) cases, the experts concluded that a picture was needed for correct classification. Including the incorrect labeled pairs that experts labeled as correct, the AI model correctly labeled 65.45\% of the MaEs. 

In Experiment 2, where the examples were grouped by topic, there were 1,100 iterations. The AI model correctly labeled 812 (73.82\%) of the iterations and incorrectly labeled 288 (26.18\%). Among the incorrectly labeled iterations, the experts found that in 121 (11.00\%) cases, the AI was mistaken, while in 111 (10.09\%) cases, the experts agreed with the AI's labeling. In 56 (5.09\%) cases, the experts determined that a picture was necessary for accurate classification. Including the incorrect labeled pairs that experts labeled as correct, the AI model correctly labeled 83.91\% of the MaEs. 

\section{Discussion}\label{sec7}

This study addresses how AI solutions can help educators identify students' misconceptions in algebra through the development and evaluation of a benchmark. Our research demonstrates that LLMs, specifically gpt4-turbo, can effectively identify student misconceptions with up to 83.9\% accuracy when constrained by topic and incorporating educator feedback. The benchmark, comprising 55 misconceptions and 220 diagnostic examples, is a bridge between research and classroom practice. Educator feedback validates the benchmark's relevance, with 80\% of surveyed teachers confirming its clarity and possible applicability. The study highlights both the potential and limitations of AI in this domain, particularly in areas like ratios and proportional thinking. These findings suggest that AI-driven tools, when used in conjunction with educators expertise, could enhance educators' ability to recognize student misconceptions.

\subsection{Educator Insights: Validation and Refinement}\label{sec7.1}

Educators' survey response (80 \% or more) indicated that they encounter these misconceptions among their students suggest that the dataset contains misconceptions that are representative to those in classrooms. The structure interviews provided insights into the relevance and potential application of the datset. Teachers' responses highlighted the need to expand the dataset to include misconceptions from other math domains such as geometry. The fact that one teacher mentioned a new misconception not found in the research emphasizes the value of incorporating educators' expertise in the dataset creation. The importance of the collaboration with teachers was also present in the feedback regarding educators' expected clarity in the answers of the misconception's examples. Teachers' feedback on the presentation of the dataset makes it clear that its formatting needs to be carefully designed to maximize its usefulness for educators.

Another important insight is the impact of technology such as Minecraft on students' understanding and the need for research to stay-up-to-date with the learning landscape. This highlights the influence of the environment in learning. A limitation of our study is the lack of comprehensive demographic data in the source research, which can compromise its applicability. Because we aim to use this dataset to serve students in educational settings with low math scores, we attempt to overcome the limitation by receiving feedback from educators serving that population. Further feedback from educators is needed, as well as bringing the benchmark's applications to be evaluated by students in contexts of low math achievement. 

\subsection{Topic-Specific Training Boosts LLM Accuracy}\label{sec7.2}

The CM (Figures \ref{confusionmatrix} \& \ref{bargraph}) and the varying performance of the model across different topics highlight another critical aspect of the challenge in accurately identifying misconceptions: the fact that the MaEs in the dataset are at different levels of abstraction. Errors and misconceptions are related but differ in meaningful ways that can affect the LLM's classification accuracy, and even between the levels of misconceptions and errors, superset and subset were found. To better understand the relationships between MaEs, we propose further research in visualizing how understanding more complex MaEs depends on grasping more basic ones. For example, to understand a function's rate of change, a learner might first need to comprehend absolute vs. relative proportionalities, which in turn requires understanding the relationship between two quantities. (See Appendix \ref{A4} for additional examples.) This hierarchical analysis can help prioritize which misconceptions to address first, potentially enabling learners to overcome multiple related misconceptions more efficiently.

Both experiments' precision and recall scores demonstrate the model's overall effectiveness in identifying misconceptions using the MaE dataset. In Experiment 1, where testing examples were randomly selected from the entire dataset, the model achieved a precision of 0.526 and a recall of 0.529 per MaE. The model's ability to correctly identify misconceptions in more than half of the cases, despite the diverse range of MaEs and the complexity of the relationships between them, highlights its potential for supporting educators in diagnosing students' misconceptions. When we include the feedback from the two expert educators in relabeling the AI model incorrect matched to correct matches the accuracy goes up to 65.45\% for experiment 1.

The higher performance of the model in the second experiment, where testing examples were randomly selected within each topic, suggests that incorporating topic-specific information during model training and evaluation could lead to improved accuracy in identifying misconceptions. When we include the feedback from the two expert educators in relabeling the AI model incorrect matched to correct matches the accuracy goes up to 83.91\% for experiment 2. This finding might support the theory that the model is more capable of matching patterns than understanding the misconception classification task. An application of this finding is that by leveraging the vast body of research on math misconceptions and organizing it by topic, AI models can better support educators in diagnosing students' misconceptions and providing targeted interventions. 
 
Additionally, future research should explore the optimal balance between topic-specificity and generalizability. This will ensure that the model can effectively identify misconceptions across various topics while benefiting from the increased accuracy provided by topic-constrained testing. Developing AI models that can better capture the underlying concepts and relationships between misconceptions, rather than relying solely on pattern matching, and training the model on a more extensive and diverse dataset that includes labeled students' wrong answers could be potential avenues for improvement. 

A limitation of this study is the exclusive use of gpt4-turbo. Future research should use a multimodal approach to address the need for further exploration of visual representations in misconception identification and include other models. This paper focuses exclusively on the examples of misconceptions and their match to a particular MaE from the dataset. In further studies, we will address the recommendation component.

A potential approach to enhance the model's performance, particularly in complex topics like ratios and proportional thinking, is to provide the model with more computing capacity and employ techniques such as chain-of-thought prompting \cite{wei2022chain}. We may observe improved performance by allowing the model to explain its reasoning before making a decision. For example, the model could be instructed to reverse engineer a student's thought process to arrive at an error and then provide the MaE prediction based on that reasoning. This approach might enable the model to capture better the nuances and underlying concepts that lead to specific misconceptions, thereby improving its overall performance. Future research should explore the effectiveness of such techniques in enhancing the model's abilities.

\subsection{Potential Applications}\label{sec7.3}

Our analysis can determine the potential misconceptions per topic, which can be useful to educators who can prepare to address them when teaching. The human expert review process further highlighted the complexity of misconceptions, as evidenced by the different types of errors identified. This complexity poses challenges for AI models and educators in accurately identifying and addressing misconceptions. Educators, particularly those with high student-to-teacher ratios or diverse classrooms, may only observe students' errors without understanding the underlying misconceptions -as an educator mention during the interview- as many misconceptions can lead to the same incorrect answer. Educators' lack of insights into learners' thinking processes connects with educators' suggestion of using the dataset in conjunction with AI as a diagnostic tool during our survey. 

Another application is that the dataset developed in this study can be used to create distractor options in multiple-choice questions \cite{feng2024exploring, mcnichols2023automated}. By designing questions targeting specific misconceptions and analyzing students' responses, educators can gain insights into the prevalent misconceptions in their classrooms and tailor their instruction accordingly, even when direct student-teacher interaction is limited. This approach demonstrates the potential for collaboration between AI researchers and domain experts in identifying areas for improvement and providing insights into the nature of misconceptions. These findings have applications for designing teaching methodologies that detect students' misconceptions, prepare teachers for detecting them, predict which possible misconceptions can be related to an algebra topic, and give AI tools and parameters to improve their misconception matching. 

\section{Conclusion}\label{sec8}

Our initial exploration give promising results towards the possibility to build practical tools in the near term to understand better where learners are in constructing mathematical thinking. Even though from a small sample, the unanimous interviewed educators' interest in using the dataset to diagnose misconceptions and learn more about student's mathematical thinking highlights the potential impact of this research on educational practices. Educators' feedback underscores the importance of developing AI tools that can provide insights into learners' math misconceptions. 

In conclusion, the results emphasize the importance of topic-constrained testing, the complexity of misconceptions, the need for multimodal approaches -as many questions and answers require graph and pictures, and the challenges in accurately identifying misconceptions using AI models. As we are working towards a dataset of research-based recommendations to overcome the MaEs, we hope our current dataset can support educators and AI platforms in meeting learners where they are with what they need to be successful in learning algebra.

\bibliography{sn-bibliography}

\appendix

\section{Appendix}
\subsection{Dataset}
\label{A1}

For each misconception, the dataset includes:
Meta Data
\begin{enumerate}
    \item ID
    \item Misconception description
    \item Topic
    \item Four research based examples
\end{enumerate}
Research-Based Examples per MaE data
\begin{itemize}
    \item Four research-based examples, each including:
\end{itemize}
\begin{itemize}
        \item A question
        \item An incorrect answer demonstrating the misconception
        \item The correct answer
\end{itemize}
\begin{itemize}
    \item When a graph is part of the question or answer the examples include:
\end{itemize}

\begin{itemize}
    	\item An example with a description of the graph.
    	\begin{figure}[H]
        	\centering
        	\includegraphics[width=1\linewidth]{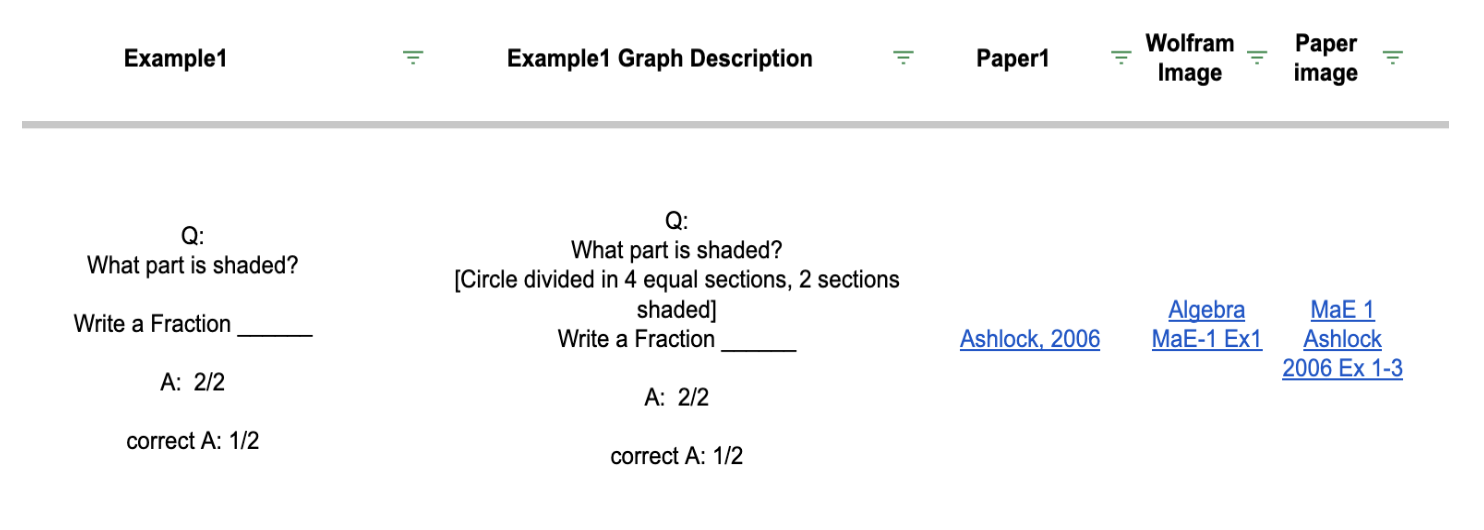}
         \caption{Example with a description of the graph}
    \label{fig:Ex1}
    	\end{figure}
    	\item A link to the image from the paper
    	\begin{figure}[H]
        	\centering
        	\includegraphics[width=0.25\linewidth]{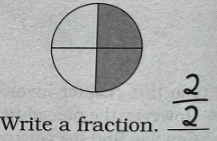}
         \caption{Example of one image from a paper}
    \label{fig:ImageAshlock}
    	\end{figure}
    	\item Reference link to the source paper for each example
\end{itemize}

\begin{figure}[H]
    \centering
    \includegraphics[width=0.65\linewidth]{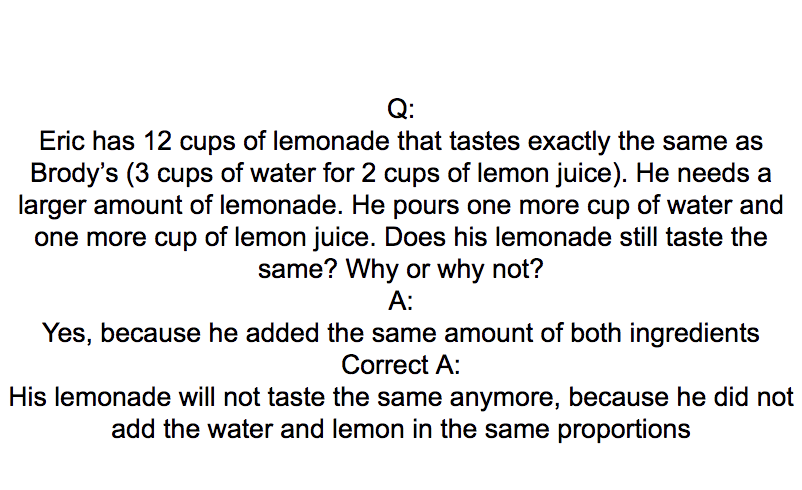}
    \caption{One example for MaE28}
    \label{fig:EricEx}
    
\end{figure}

\begin{figure}[H]
    \centering
    \includegraphics[width=1\linewidth]{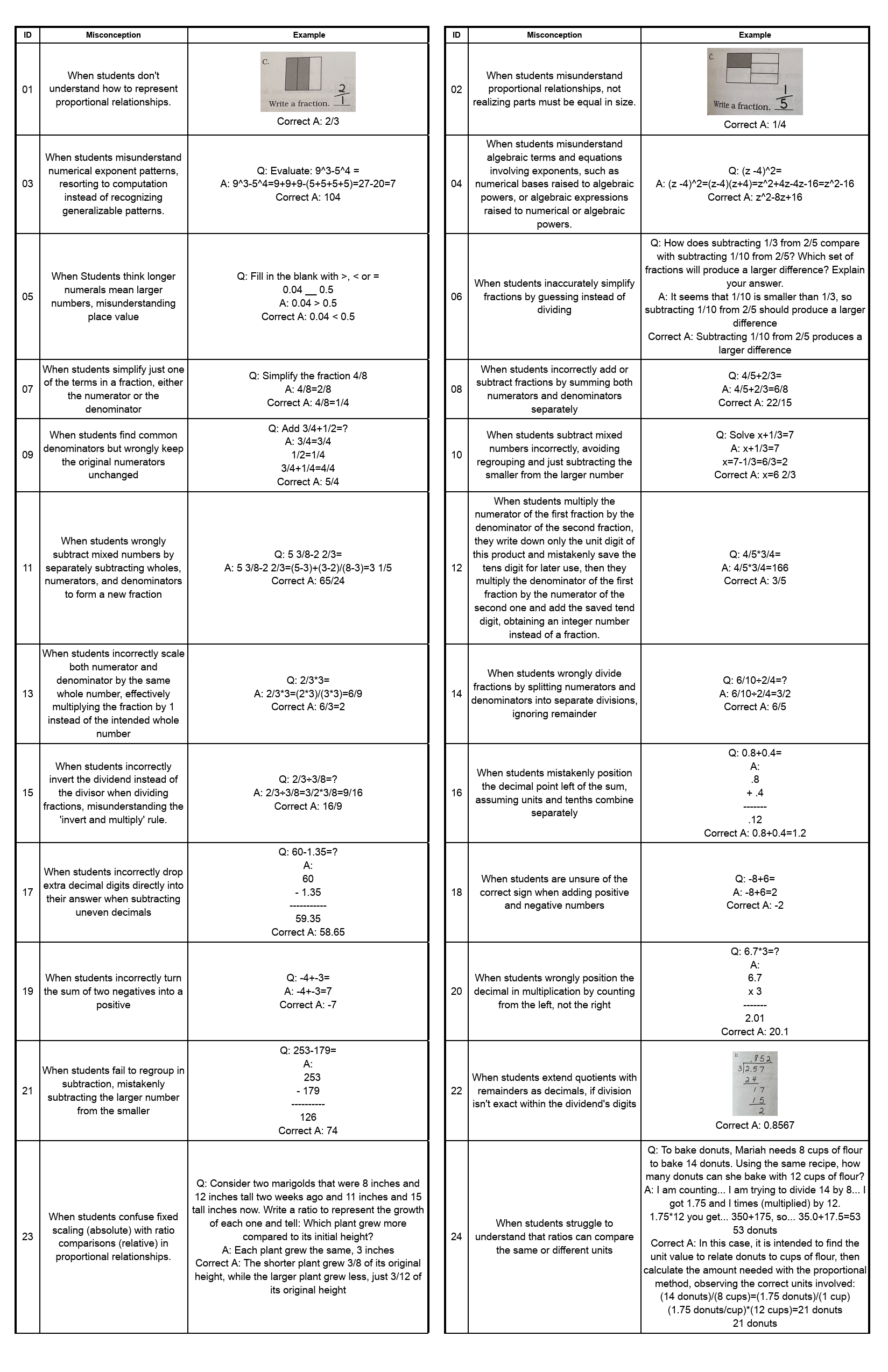}
    \caption{Dataset with one example. MaEs 1-24}
    \label{fig:BTE1}
\end{figure}

\begin{figure}[H]
    \centering
    \includegraphics[width=1\linewidth]{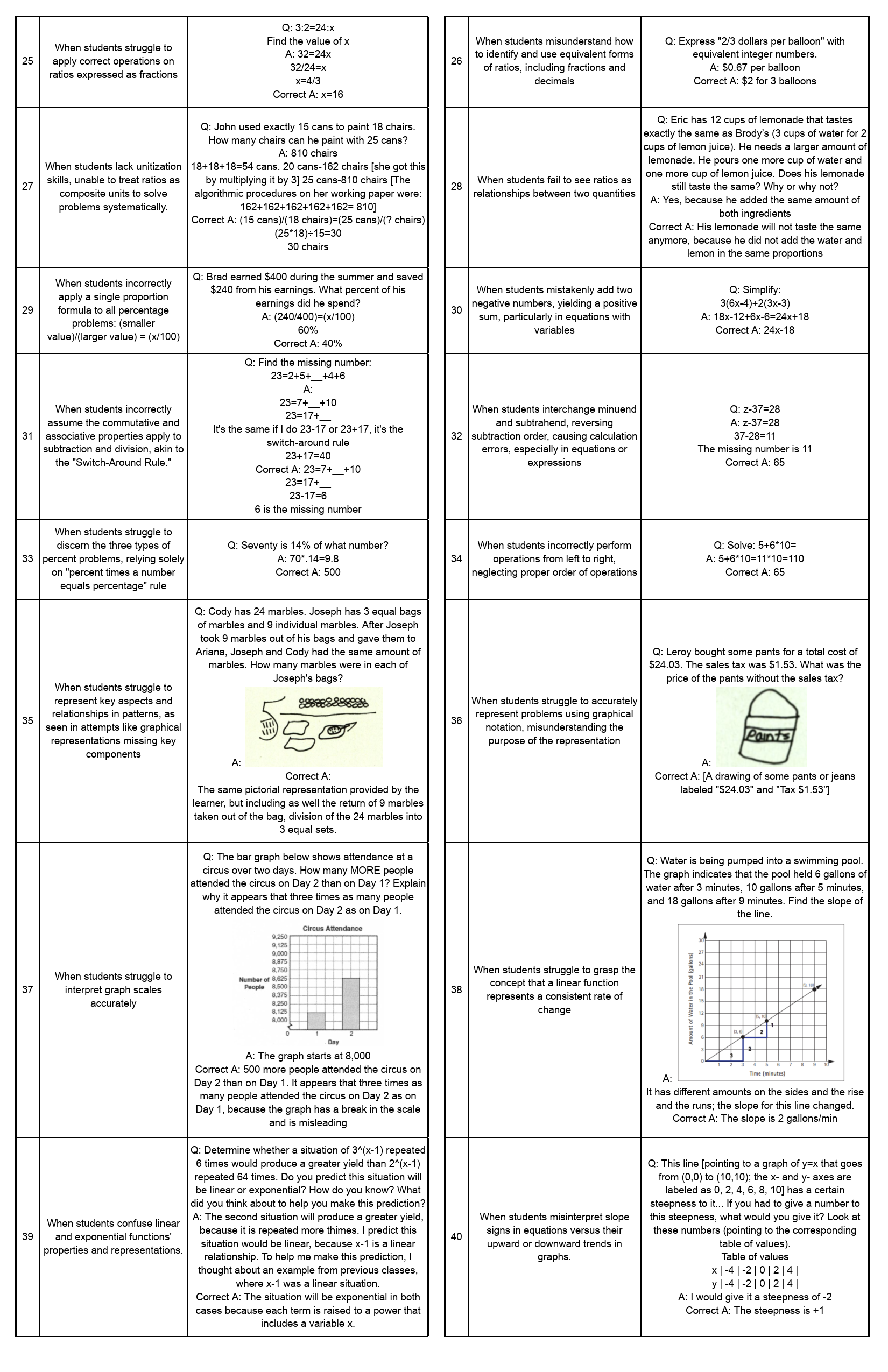}
    \caption{Dataset with one example. MaEs 25-40}
    \label{fig:BTE2}
\end{figure}

\begin{figure}[H]
    \centering
    \includegraphics[width=1\linewidth]{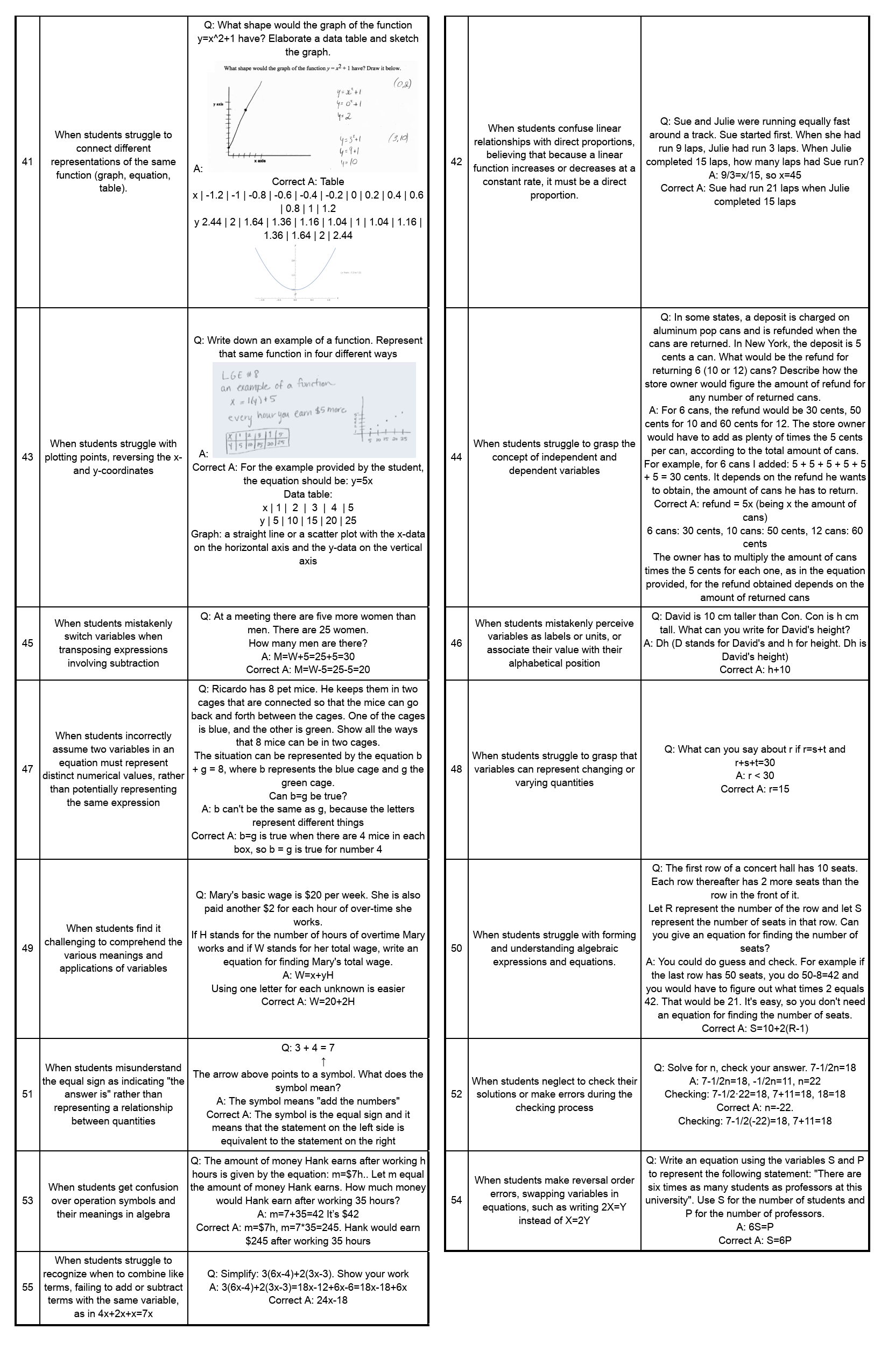}
    \caption{Dataset with one example. MaEs 41-55}
    \label{fig:BTE3}
\end{figure}

\subsection{Prompt}
\label{A2}

\textbf{The prompt for the experiments was as follows
}

Prompt: 

You are an expert tutor on middle school math with years of experience understanding students' most common math mistakes.
You have identified a set of common mistakes called Misconceptions, and you use them to diagnose student's answers to math questions.
You have also developed a labeled dataset of question items, and diagnosed them with the appropriate misconception id.
Using the set of misconceptions and the labeled dataset, your task today is to take some items of unlabeled data and provide a diagnosis for each unlabeled item.

Here is the list of misconceptions together with a brief description:
{description}

Here is  dataset of question items already labeled with a diagnosis:
\{train\_example\}

Here is the unlabeled dataset of question items. For each question item below, give a diagnosis.
\{items\_test\}

FOR EACH UNLABELED ITEM, OUTPUT A LINE WITH THE FOLLOWING FORMAT:
UNLABELED ITEM \$N DIAGNOSIS: MaE\_ID

DO NOT INCLUDE ANY ADDITIONAL OUTPUT.

\subsection{Tables with examples of MaEs}
\label{A3}

\begin{figure}[H]
    \centering
    \includegraphics[width=1\linewidth]{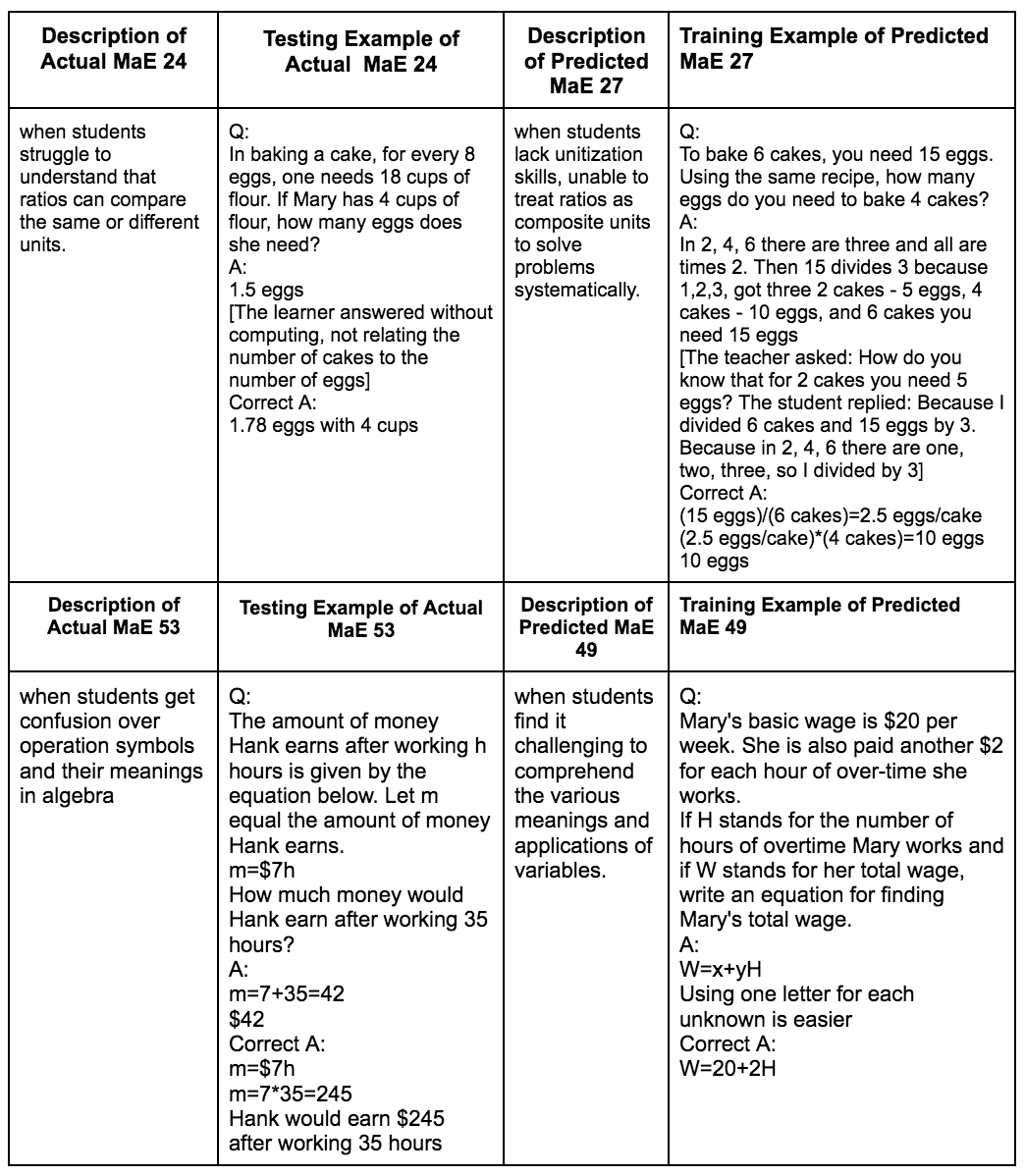}
    \caption{Example of experts agreeing with AI mistaken label per topic}
    \label{Table5.1}
\end{figure}

\begin{figure}[H]
    \centering
    \includegraphics[width=1\linewidth]{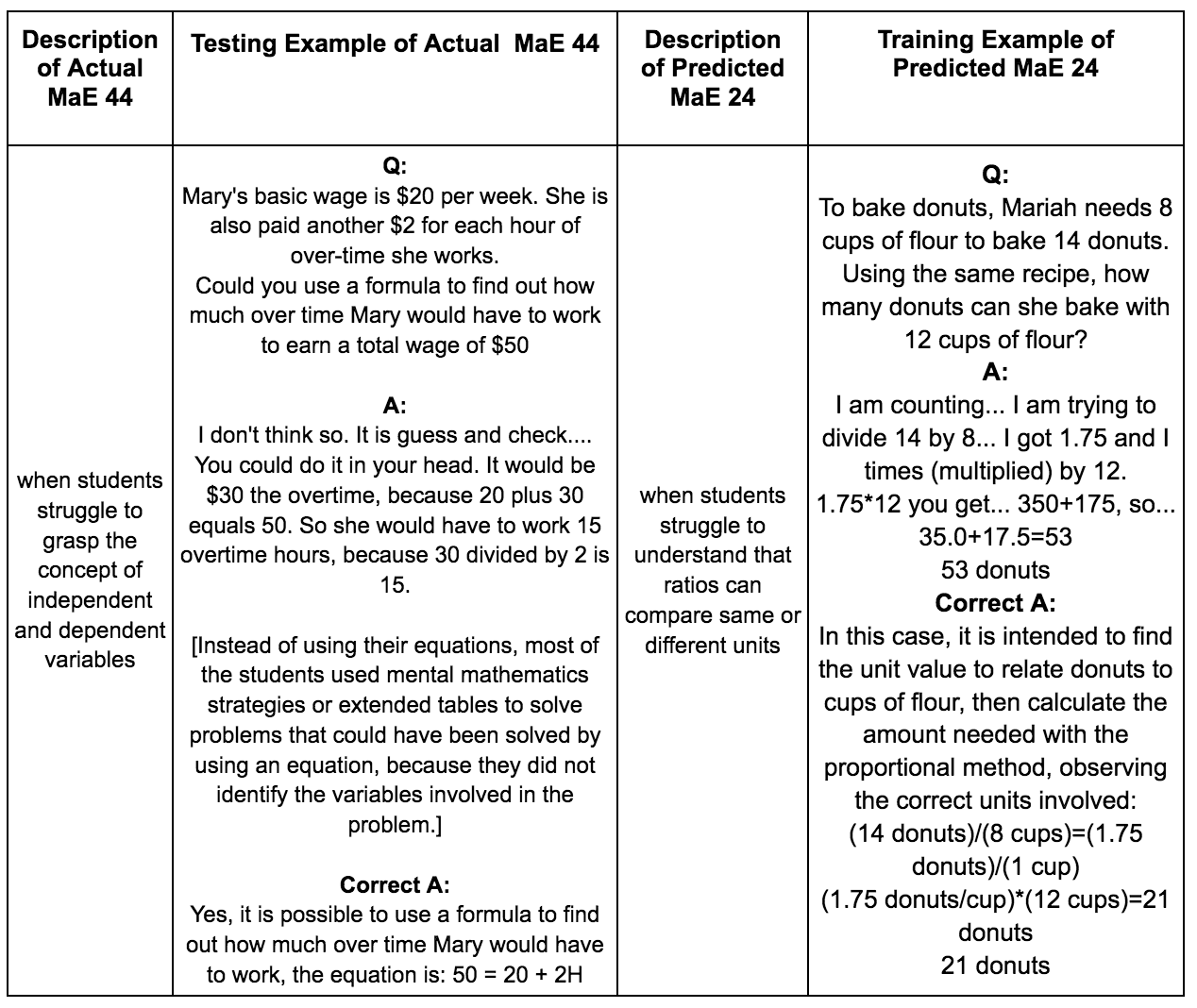}
    \caption{Example of experts agreeing with AI mistaken label in different topics}
    \label{Table5.2}
\end{figure}

\begin{figure}[H]
    \centering
    \includegraphics[width=1\linewidth]{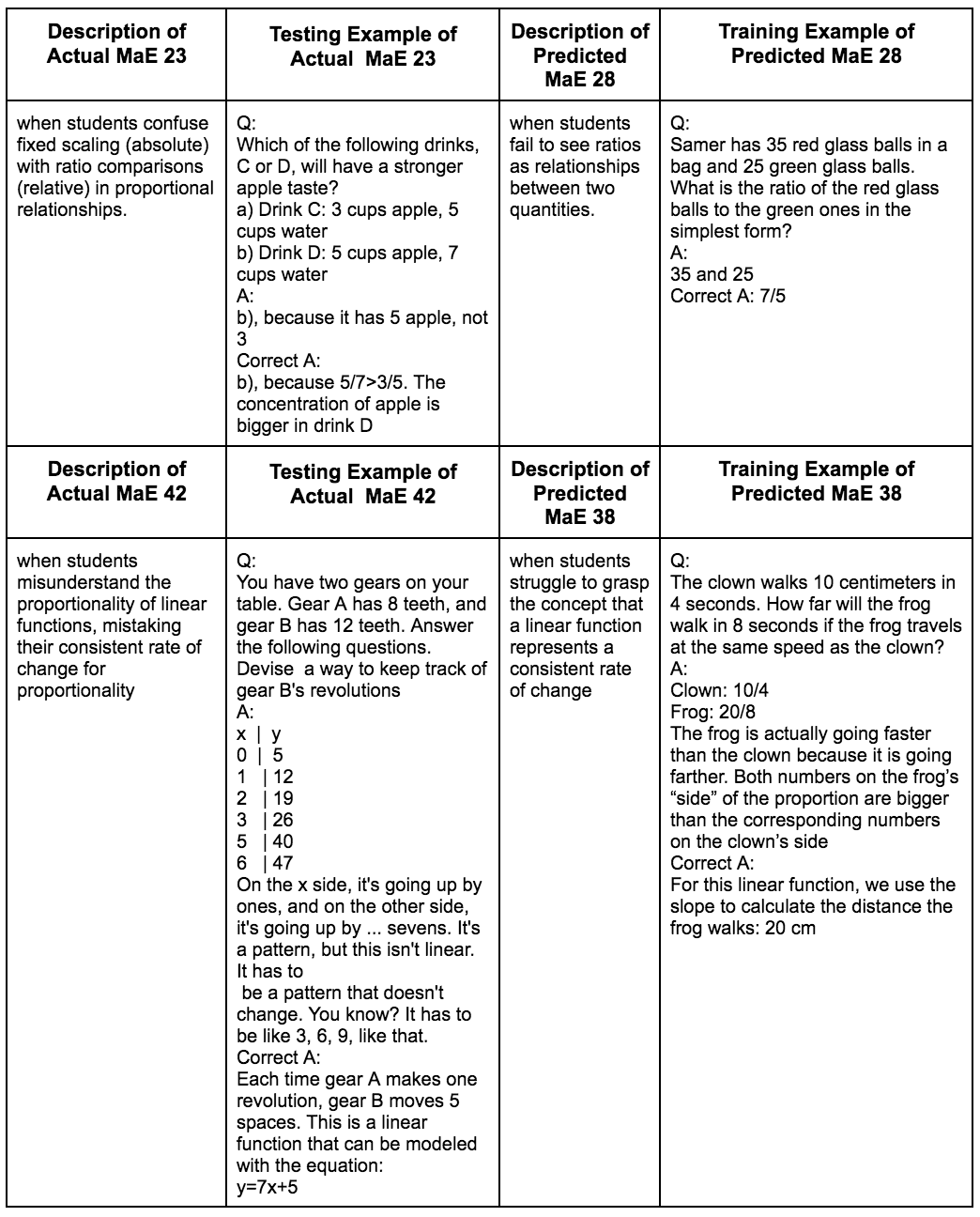}
    \caption{Examples of a MaE being a subset or superset of another MaE}
    \label{Table6}
\end{figure}
\begin{figure}[H]
    \centering
    \includegraphics[width=1\linewidth]{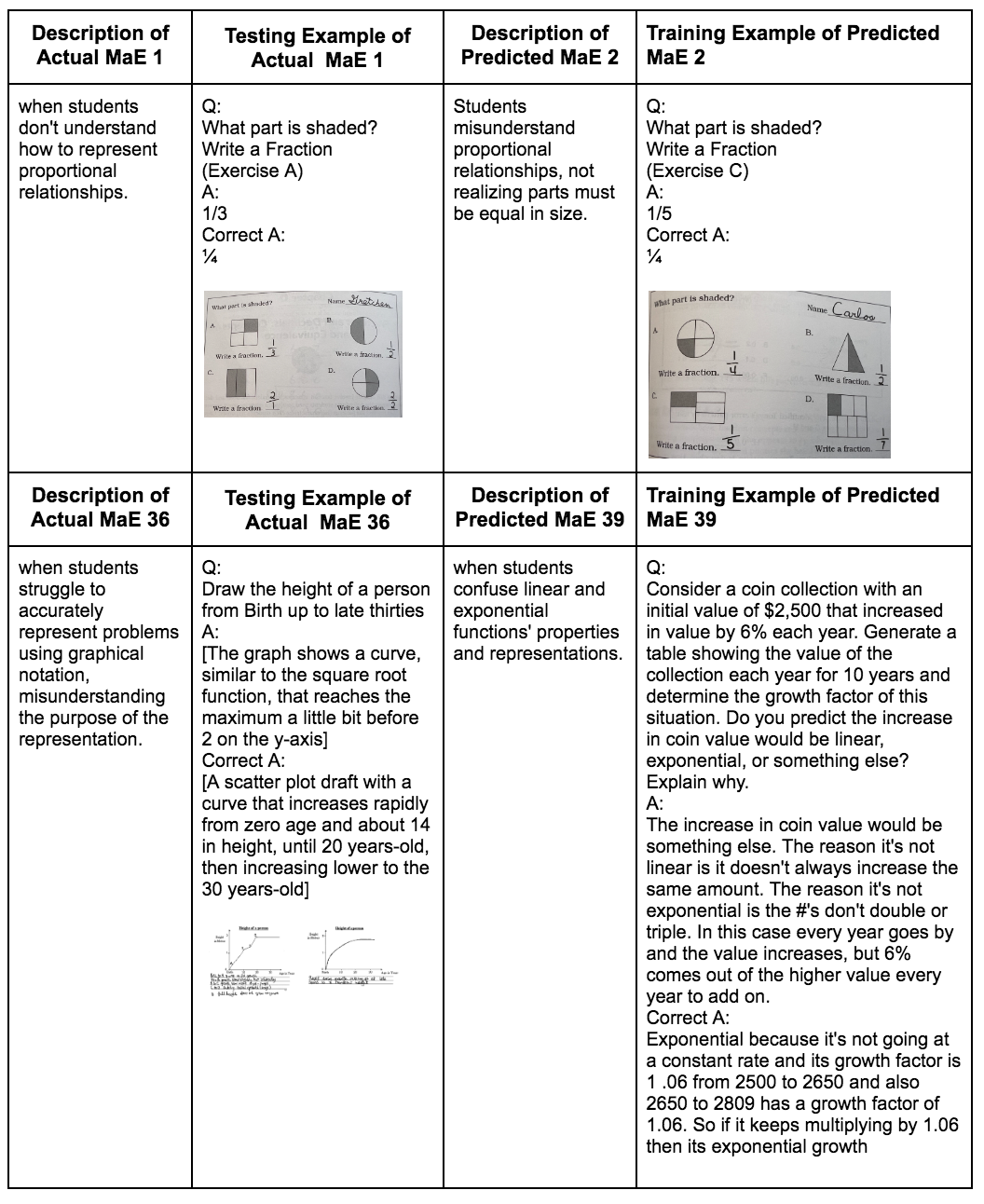}
    \caption{An example of a MaE that contains a picture}
    \label{Table7}
\end{figure}
\begin{figure}[H]
    \centering
    \includegraphics[width=1\linewidth]{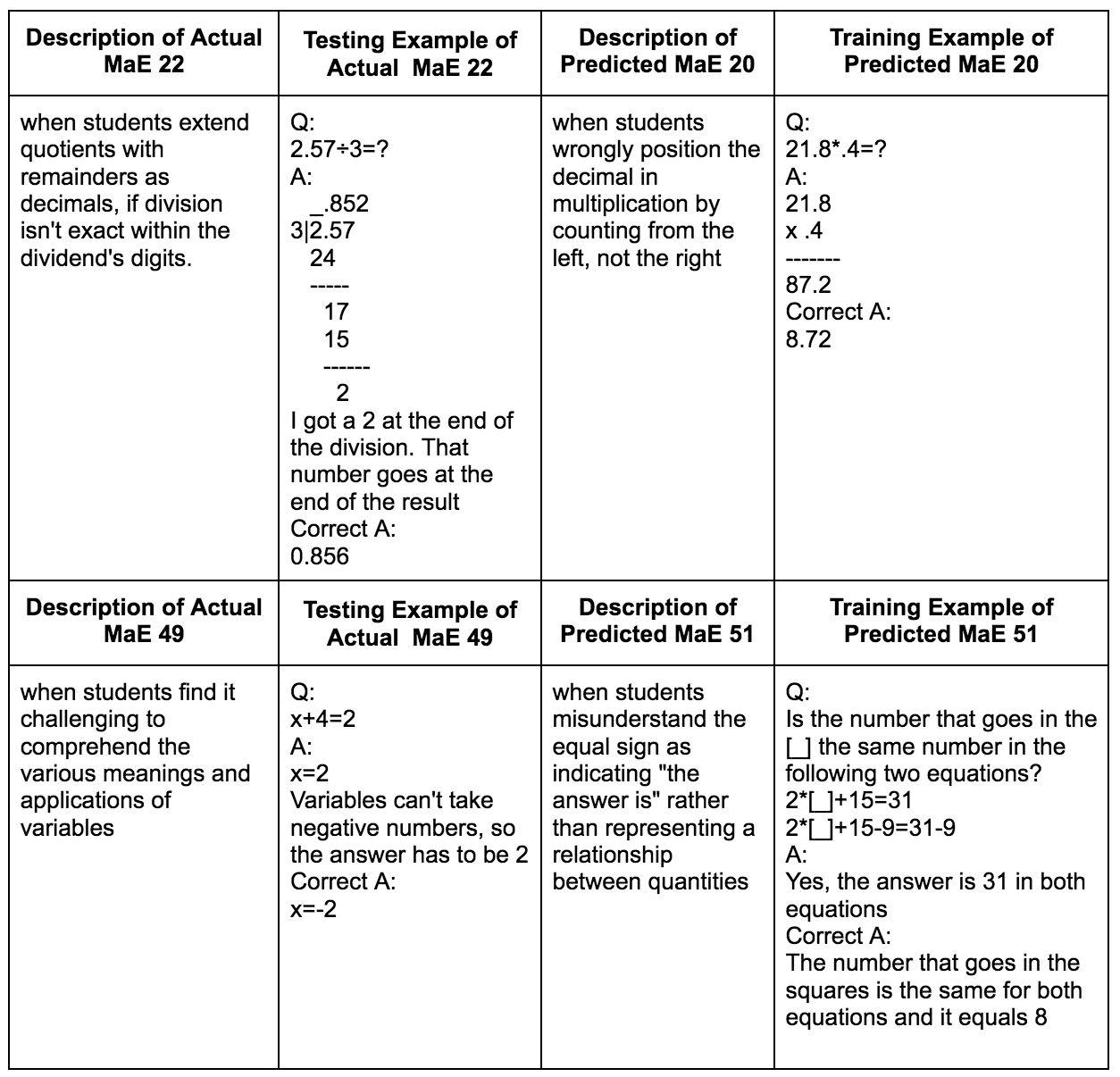}
    \caption{An example of a MaE where AI was mistaken}
    \label{Table8}
\end{figure}

\subsection{Example of relationship between misconceptions}
\label{A4}

\begin{figure}[H]
    \centering
    \includegraphics[width=1.2\linewidth]{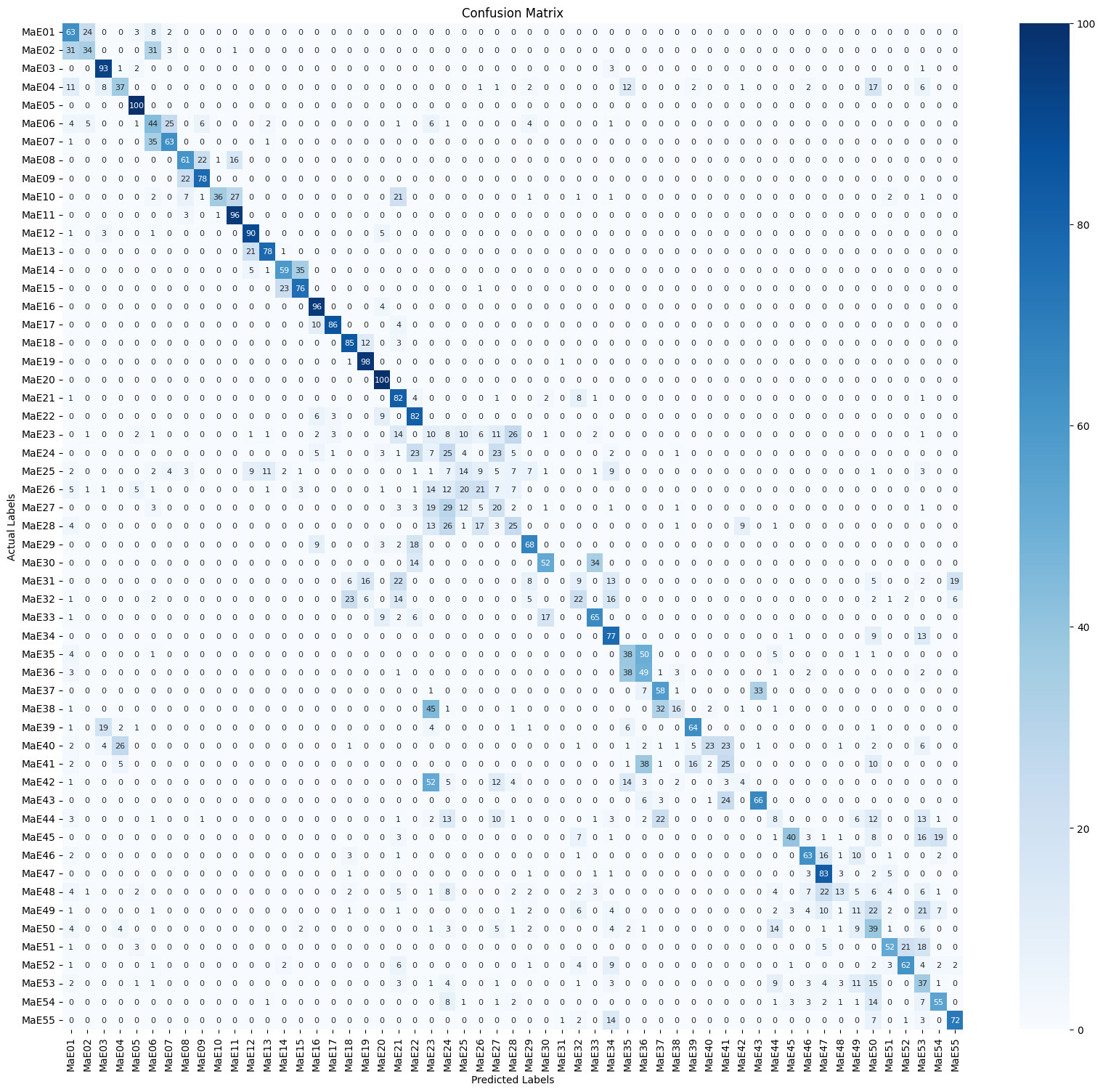}
    \caption{Confusion Matrix (CM)}
    \label{confusionmatrix}
\end{figure}

The CM reveals patterns in the model's performance: MaEs from Patterns, relationships, and functions (PRF) can be classified as Ratios and proportions (RP) MaEs, but not vice versa. An explanation can be that RP does not contain concepts from PRF, but PRF does. Expert educators agreed that is possible given that MaEs 38 and 42 are about functions' rate of change and MaE 23 is \textit{"When students confuse fixed scaling (absolute) with ratio comparisons (relative) in a proportional relationship,"} . See details on this example in Appendix \ref{A4}. Applying this pattern shows the MaEs that are needed to solve problems for other MaEs.

Expert educators agreed that is possible given that MaEs 38 and 42 are about functions' rate of change and MaE 23 is \textit{"When students confuse fixed scaling (absolute) with ratio comparisons (relative) in a proportional relationship,"} MaE 23 example-4:

\textit{Question: Consider two marigolds that were 8 inches and 12 inches tall two weeks ago and 11 inches and 15 tall inches now. Write a ratio to represent the growth of each one and tell: Which plant grew more? 
Answer: Each plant grew the same, 3 inches 
Correct Answer: The shorter plant grew 3/8 of its original height, while the larger plant grew less, just 3/12 of its original height.}

Furthermore, by following that logic, the CM shows an additional level of possible misunderstanding now connected with MaE 23, MaE 28 "when students fail to see ratios as relationships between two quantities", MaE 28 example-4:

\textit{Question: Eric has 12 cups of lemonade that tastes exactly the same as Brody's (3 cups of water for 2 cups of lemon juice). He needs a larger amount of lemonade. He pours one more cup of water and one more cup of lemon juice. Does his lemonade still taste the same? Why or why not?
Answer: Yes, because he added the same amount of both ingredients
Correct Answer: His lemonade will not taste the same anymore, because he did not add the water and lemon in the same proportions.}

\end{document}